\documentclass[sigconf]{acmart}
\usepackage{soul}
\usepackage{color}
\usepackage{graphicx}
\usepackage[acronym,shortcuts]{glossaries}
\usepackage{algorithm}
\usepackage{algpseudocode}
\usepackage{balance}
\usepackage{subcaption}
\usepackage{comment}
\newacronym{b5g}{B5G}{Beyond Fifth Generation}
\newacronym{6g}{6G}{Sixth Generation}
\newacronym{drl}{DRL}{Deep Reinforcement Learning}
\newacronym{iot}{IoT}{Internet of Things}
\newacronym{mab}{MAB}{Multi-Armed Bandit}
\newacronym{pdr}{PDR}{Packet Delivery Ratio}
\newacronym{per}{PER}{Packet Error Rate}
\newacronym{rl}{RL}{Reinforcement Learning}
\newacronym{rssi}{RSSI}{Received Signal Strength Indicator}
\newacronym{sinr}{SINR}{Signal to Interference plus Noise Ratio}

\newcommand{\ale}[1]{\textcolor{black}{#1}}
\newcommand{\emi}[1]{\textcolor{black}{#1}}
\newcommand{\emiAres}[1]{\textcolor{black}{#1}}

\setcopyright{acmcopyright}
\copyrightyear{2018}
\acmYear{2018}
\acmDOI{10.1145/1122445.1122456}

\acmConference[]{}{}{}
\acmBooktitle{}
\acmPrice{}
\acmISBN{}

\begin{document}

\title{%POULPE: 
FOLPETTI: 
A Novel Multi-Armed Bandit Smart Attack \\ for Wireless Networks
}

\author{Bout Emilie}
\email{emilie.bout@inria.com}
\affiliation{%
  \institution{Inria-Lille}
  \city{Lille}
  \country{France}
  \postcode{59650}
}

\author{Brighente Alessandro}
\email{alessandro.brighente@unipd.it}
\affiliation{%
  \institution{Tsinghua University}
  \streetaddress{30 Shuangqing Rd}
  \city{Padova}
  \country{Italy}
  \postcode{35100}}

\author{Mauro Conti}
\email{mauro.conti@unipd.it}
\affiliation{%
  \institution{University of Padova}
  \city{Padova}
  \country{Italy}
  \postcode{35100}}

\author{Valeria Loscri}
\email{valeria.loscri@inria.com}
\affiliation{%
  \institution{Inria-Lille}
  \city{Lille}
  \country{France}
  \postcode{59650}
}
\begin{abstract}
 %\hl{at most 9 pages}
%  Due to the popularity and massive diffusion of wireless technology, jamming attacks represent one of the most important issues in wireless systems. This type of attack can generate denial-of-service effects, by sending malicious data causing intentional interference. Channel hopping has been adopted as effective defense method against such a type of attack. Some advanced technique has been proposed in this direction, for letting the transmitter-receiver pair to be re-synchronized on a new channel, without leaking the information to the attacker. 
%The increasing number of \ac{iot} nodes supported by \ac{b5g} and \ac{6g} networks provides a large attack surface. In particular, attackers may exploit the openness of the wireless channel to launch different types of attacks. To guarantee the service availability upon detection of an attack, devices can dinamically select a non-attacked channel via channel hopping according to a predefined pattern.
\ale{Channel hopping provides a defense mechanism against jamming attacks in large scale \ac{iot} networks.} However, a sufficiently powerful attacker may be able to learn \ale{the channel hopping pattern and efficiently predict the channel to jam.} 
%In particular, recent works proposed to use reinforcement learning to learn the victim's channel hopping pattern. 
%In addition, \ale{since the victim may select a new hopping pattern upon attack detection, literature solutions require the attacker to periodically learn the victim's pattern}, therefore resulting in a non-continuous attack.

In this paper, we present FOLPETTI, a \ac{mab}-based attack to dynamically follow the victim's channel selection in real-time. Compared to previous attacks implemented via \ac{drl}, FOLPETTI does not require recurrent training phases to capture the victim's behavior, allowing hence a continuous attack. We assess the validity of FOLPETTI by implementing it to launch a jamming attack. We evaluate its performance against a victim performing random channel selection and a victim implementing a \ac{mab} defence strategy. We assume that the victim detects an attack when more than $20\%$ of the transmitted packets are not received, therefore this represents the limit for the attack to be stealthy. In this scenario, FOLPETTI achieves a $15\%$ success rate for the victim's random channel selection strategy, close to the $17.5\%$ obtained with a genie-aided approach.\emi{ Conversely, the DRL-based approach reaches a success rate of $12.5\%$, which is $5.5\%$ less than FOLPETTI.} \emi{We also confirm the results by confronting FOLPETTI with a \ac{mab} based channel hopping method.} Finally, we show that FOLPETTI creates an additional energy demand independently from its success rate, therefore decreasing the lifetime of \ac{iot} devices. 

\end{abstract}




\maketitle
\glsresetall
\section{Introduction}
Guaranteeing security in \ac{iot} represents a challenging task due to the large number of connected devices and their largely distributed nature~\cite{hassan2019current}. \ale{The attack surface of this technology is expected to further expand, due to the introduction of massive \ac{iot} networks}~\cite{guo2021enabling}. Furthermore, these devices are mostly connected via wireless communications, whose openness represents an additional security challenge. \ale{A common strategy to mitigate attacks is to move the communication from the attacked channel to a free one.} \ale{Each user is hence} given a set of different channels \ale{to use} for communication. Channel hopping is the process by means of which a user selects, upon detecting an attack or a communication failure, a new channel for transmission. This is particularly \ale{useful} in attacks such as jamming, where a malicious user intentionally degrades the quality of the victim's channel \ale{preventing the} successful delivery of packets \cite{grover2014jamming}. \ale{In this case, the victim selects an alternative free channel according to a predefined strategy. }

Most of the research \ale{on channel hopping} focused on the design of strategies to optimally select the new channel. In particular, given different attackers capabilities, researchers focused on the design of the best channel hopping strategy to guarantee continuous service availability. Relevant examples include sequential channel hopping (i.e., move to the next available channel), random channel hopping (i.e., select a random channel from those available)~\cite{lee2010randomized}, game theory approaches~\cite{tang2018jamming, namvar2016jamming}, and machine learning approaches~\cite{upadhyaya2019machine,han2017two,xiao2018anti}. However, given the increasing availability of cheap smart and powerful devices, an attacker \ale{can design an optimal strategy too.} In particular, given a certain defence strategy, the attacker \ale{can} learn the victim's channel hopping pattern to \ale{predict the channel to }jam. To this aim, an attacker might use \ac{drl}, where she first observes the victim's behavior, and then attack according to the inferred channel hopping pattern~\cite{zhong2020adversarial}. \ale{In response, the victim selects a new channel hopping strategy to reduce the effectiveness of the attacker}. In this case, the attacker needs to periodically interrupt the attack to learn the new \ale{victim's} channel hopping pattern. The  \ac{drl} strategy \ale{also} relies on the assumption that, during the learning phase, the victim's channel hopping strategy remains static. \ale{Therefore, learning-based methods do not represent an efficient solution against channel hopping. Based on these observations, an effective attacks strategy does not need to rely on a particular victim's behavior, and should provide attack continuity.}

In this paper, we propose FOLPETTI, a novel attack strategy against channel hopping that guarantees the continuity of the attack. %Thanks to FOLPETTI, the attacker does not need to periodically interrupt the attack to learn the victim's channel hopping strategy. 
We exploit a \ac{mab} approach to resolve the exploit-explore dilemma at the attackers' side. Thanks to our online approach, we guarantee both the attack continuity and the independence from the assumption of a victim's static channel hopping strategy. As a relevant example, we test FOLPETTI in a jamming scenario considering two defence strategies: i) random channel hopping, and ii) the smart channel hopping strategy implemented in~\cite{C-4}. \emi{We compare FOLPETTI against three other attacking strategies: i) Random strategy, where the attacker randomly selects a channel to attack; ii) \ac{drl} strategy, where the attacker uses \ac{drl} to predict the channel to attack; and iii) Optimal strategy, i.e., a genie aided method.}  We assume that the victim detects an attacker when the percentage of correctly delivered packets falls below $80\%$. This hence represents the limit that the attacker needs to achieve to be stealthy. Our results show that, in case of random channel hopping, FOLPETTI reduces the \ac{pdr} to $82\%$, close to the $80\%$ \ac{pdr} attained with a genie aided method (optimal solution). \emi{The FOLPETTI attack is continuous, therefore causing an higher impact on the network compared to that of the \ac{drl}-based attack, which needs to periodically stop and learn the best strategy. Indeed, we show that the \ac{drl} based attack reduces the \ac{pdr} to $85\%$, i.e  $5\%$ less than FOLPETTI.} Also when the transmitter adopts the smart channel hopping strategy, we demonstrate that FOLPETTI is able to reduce the \ac{pdr} more significantly compared to other methods. Moreover, we prove that FOLPETTI significantly increases the number of retransmissions and consequently the energy demand of the victims. 
%We show that, due to \ale{the increased number of} retransmissions, FOLPETTI creates a $3$J additional energy demand when attacking a random channel hopper, and an additional $1$J energy demand when attacking a smart victim. %\hl{review later}
%The previous smart attacks relying on Machine Learning (ML) approaches, need to interrupt the attack functionality for learning/exploration purpose, with a reduced effects on the attack impact on the network.

The main contributions of the paper can be summarized as follows.
\begin{itemize}
    \item We design FOLPETTI, a new smart attack strategy against channel hopping. FOLPETTI is based on a \ac{mab} approach able to perform an optimal channel selection.
    \item We integrate the FOLPETTI strategy with a jamming attack and demonstrate that the attacker is able to learn the victim's channel hopping strategy and attack online. This guarantees a continuous attack.
    \item We show that FOLPETTI increases the number of retransmissions and hence increases the nodes' energy demand, shortening devices lifetime.
    \item \emi{We develop a jamming module with the latest version of discrete event simulator NS-3 (Network Simulator-3). We add some essential needs for simulating modern jamming attacks, such as the machine learning-based. }
\end{itemize}

The rest of the paper is organized as follows. In Section~\ref{relwork} we present the representative works on attacks against channel hopping. In Section~\ref{method} we detail the FOLPETTI strategy and how this approach can be combined with a jamming attack. Section~\ref{section-mitigation} describes the approach used for detecting jamming attacks and the channel hopping strategy implemented after an attack detection. Section~\ref{performance} provides an analysis of the behavior of the FOLPETTI strategy. \ale{Then, in Section~\ref{sec:evaluation} we compare the performance of FOLPETTI with those of other relevant attacks \emi{in two different scenarios}. In section~\ref{discussion} we provide some insights on the implementation of such a kind of attack in a wireless communication network. We discuss in Section~\ref{sec:countermeasures}  possible countermeasures to prevent FOLPETTI.} Finally, we conclude the paper and provide some future research direction in Section~\ref{concl}. 

\section{Related Literature}\label{relwork}
In this section we review the literature \ale{discussing} channel hopping schemes. \ale{We first discuss the defence perspective, to gain insights on how victims select their hopping strategies. Then we discuss the attack perspective to clearly show the novelty of our attack strategy.}

\textbf{Defence Perspective.} 
Channel hopping \ale{solutions} range from the most naive (i.e., sequential channel hopping) to the most elaborated and smart (e.g., via \ac{rl}). \ale{We focus on the second class, and in particular on \ac{rl} where devices can decide on the next channel based on the positive and negative feedback received from the network.} The authors in~\cite{han2017two} propose a framework where, in case of heavy jamming, a user might leave the network and connect to another access point. The \ac{rl} framework exploits the past observations to decide whether to leave the network or select a new channel. A similar \ac{rl}-based mobility concept has been exploited by the authors of~\cite{xiao2018anti}  to decide whether a user should combat the jammer or leave. Both~\cite{han2017two} and~\cite{xiao2018anti} used quantized \ac{sinr} values as \ac{rl} states, \ale{therefore being limited in real world scenarios}. To deal with the problem of infinite \ac{sinr} states, the authors in~\cite{liu2018anti} proposed to use the spectrum waterfall as \ac{rl} states. Furthermore, their solution does not require the knowledge of the jamming pattern, and extends to the case where the attacker implements an intelligent and dynamic jamming strategy. The authors in~\cite{wang2020mean} proposed a multi-agent \ac{rl} framework to account for both the jamming attacker and the mutual interference caused by multiple legitimate nodes. A similar concept has been proposed by the authors in~\cite{zhou2021intelligent} to cope with the increasing number of devices in ultra-dense networks. All these works however did not account for the power-limited capabilities of \ac{iot} devices. The authors in~\cite{xu2020intelligent} proposed a Markov decision process-based \ac{rl} framework, showing the power consumption of their framework against three types of jammers: sweep, random, and sensing-based. However, they did not consider the effects of an intelligent jamming strategy. 

\textbf{Attack Perspective:} 
As demonstrated by the fervent literature described above, most of the literature focuses on channel hopping solutions for the mitigation.  In this paper, we focus on the attacker side and we hence review available proposals for smart jamming strategies. The authors in \cite{noori2020jamming} proposed a game-theoretic approach, where both the attacker and the victim choose the channel to hop solving for the optimal strategy. However, authors resolved to Q-learning to \ale{compute the solution of the game}, hence requiring periodical learning phases. The authors in~\cite{zhong2020adversarial} proposed a deep \ac{rl} framework, where the attacker observes the victim activity to learn the activity pattern. The attacker employs an actor-critic neural network model, where the actor tests the channel and reports result to the critic which decides on the next action. Authors assume that the victim pattern is static during the learning phase. \ale{An extension to simultaneously attacks multiple channels has been proposed} in~\cite{wang2021adversarial}, where the authors consider multiple dynamic channels. Their approach is based on a Deep Q-Network (DQN) approach with the main objective to reduce the sum-rate of the victim node. In order to implement an effective attack, the authors foresee a listening phase where the attack is in stand-by. 

\ale{Unlike the previous attacks, our solutions is able to attack in a continuous fashion, without requiring a listening phases. We hence provide a novel and efficient attacks strategy.}

\vspace{-0.50cm}

\section{Methodology}\label{method}

\ale{In this section, we describe our attack methodology. We first describe FOLPETTI,} our novel attack strategy in Section~\ref{sec:folpetti}. Then, as a relevant example, we show its implementation for a jamming attack in Section~\ref{sec:folpettiJamming}.

\subsection{FOLPETTI: a  Novel Attack Framework}\label{sec:folpetti}
%Diverse channel hopping approaches have been developed to prevent interference and possible attacks like jamming or flooding \cite{C-1,C-2}. These solutions possess the crucial advantage of not suspending the communication between the nodes of the network. In addition, channel hopping is increasingly used in other situations to extend network performance, such as in cognitive radio \cite{C-3}. 
%In the context of channel hopping, we developed a novel type of attack able to follow the victim's channel hopping pattern. This allows for a constant reduction of the network's transmission capacity. 
\ale{An efficient attacks against channel hopping has two fundamental requirements: i) it should not depend on specific assumptions on the victim's hopping pattern, and ii) it needs to be continuous in time. We here explain how FOLPETTI satisfies both these requirements.}

\ale{In FOLPETTI, the attacker follows the victim's channel selection using a \ac{mab} framework. Several \ac{mab} algorithms have been implemented at the defence side,} where the goal is to find the optimal non-attacked communication channel~\cite{C-4,C-5}. We adapted the framework of these solutions to \ale{implement an attack strategy}.
\ale{To select the optimal channel, i.e., that where the attack will have the most effect, the attacker needs to solve} the Exploit-Explore dilemma. \ale{This problem can be modeled with via \ac{mab} as a Markov Decision Process (MDP).} The MDP can be described via five tuples \(\langle S, A, P, R,\gamma \rangle \), where:
\begin{itemize}
\item  \(S\) is a finite set of states \(s\);
\item  \(A\) is a finite set of actions \(a\);
\item \(P_{a}(s^n,s^{n+1})\) is the probability that an action \(a\) in state \(s^n\) in time \(n+1\);
\item \(R_{a}(s^n,s^{n+1})\) is the expected immediate reward received after transitioning state \(s^n\) to state \(s^{n+1}\) due to action \(a\);
\item \(\gamma \in [0,1]\) is a discount factor.
\end{itemize}

The main objective of a MDP is to find a policy \(\pi\) that associates an action \ale{to} each state \(\pi:S \rightarrow{A}\) to maximize the reward. Therefore, the agent tries to maximize his reward by selecting the optimal channel. \ale{At} each time instant $n$, the attacker may stay in the previously selected state $s_i$, or move to another state $s_j$. Therefore, we define the possible actions as the set $S={s_1,s_2,...,s_S}$ of the available channels. We assume that the reward is one  \(R_{a}(s^n,s^{n+1})=1\) whenever the newly selected channel is \ale{used by a victim}. Otherwise, the reward is zero \(R_{a}(s^n,s^{n+1})=0\). 

To \ale{solve} this online decision problem, inspired by the work in \cite{C-4}, we apply the Thompson sampling algorithm as a policy. The prior distribution beta${(\alpha_j,\beta_j)}$ for each access trial is a Beta
distribution with parameters ${\alpha_j}$ and ${\beta_j}$. \ale{We denote as ${\mu_j}$ the event where the attack is successful ($R_{a} =1$). For action $s$, the probability of success is given by}
\begin{equation}
    p_j(\mu_j|S_j) = \frac{\Gamma(\alpha_j + \beta_j)}{\Gamma(\alpha_j)\Gamma(\beta_j)}(\mu_j)^{\mu_j-1}(1-\mu_j)^{\beta_j-1};
\end{equation}
where ${\Gamma(\cdot)}$ is the gamma function. If the state $s$ is selected in round t and returns a reward $R_{a}$, the prior distribution for the mean reward of arm $s$ can be updated via the Bayes rule. By utilizing the conjugacy properties, the posterior distribution for the mean reward of each arm is also a beta distribution with parameters updated based on the following rules~\cite{R-1}:

\begin{equation}
    (\alpha_s,\beta_s) \leftarrow 
    \begin{cases}
        (\alpha_s,\beta_s) & \mbox{if } a_t \neq s; \\ 
        (\alpha_s +R_{a},\beta_s +1 -R_{a}) & \mbox{if } a_t = s. 
    \end{cases}
\end{equation}

\subsection{FOLPETTI Combined with Jamming Attack}\label{sec:folpettiJamming}
To validate the effectiveness of FOLPETTI, we show its application to a jamming attack. \ale{Jamming attacks have the purpose of causing a denial of service by degrading the channel's quality and preventing the exchange of packets between legitimate nodes in the network.} The jammer has hence the option of voluntarily occupying the channel or causing collisions to corrupt the packet and force the node to retransmit. 
\begin{comment}
The effectiveness of a jamming attack is based on many parameters such as the transmission properties (e.g., modulation, power) or the characteristics of the network (e.g., routing, medium).
\end{comment}
%Consequently, by successfully predicting the communication channel selection pattern of a network, an attacker can reduce the probability of being detected. 
\ale{Furthermore, the attacker needs to act in a stealthy fashion, to avoid being detected.}

In this work, we implement a constant jamming attack to continuously jam \ale{the victim's channels}. Based on FOLPETTI model, the attacker \ale{follows the victim's channel selection and jams them as described in Algorithm~\ref{alg:cap}. Notice that in this work we assume that the attacker can jam a single channel per time instant. We will consider an attack jamming over multiple channels in future works.}

\begin{algorithm}
\caption{FOLPETTI Algorithm}
\label{alg:cap}
\begin{algorithmic}
\Require $j: \text{channel index}, c: \text{total number of channel accesses},tj : \text{number of successful transmissions so far}  $
\State $\alpha_j = \beta_j = 1$
\State $t_j = c = 0$
\While{$\text{True }$}
\For{\texttt{all j }}
        \State \text{sample} $r_j \sim $ \text{beta} $(\alpha_j +t_j,\beta_j + t_j )$
      \EndFor\\
    \text{m } = \text{ argmax }${\{\bar{r}_j\}}$;\\
    c ++ ;
    \text{JAM();}
    \If{\text{channel is occupied}}
        $t_m+ = 1$
    \EndIf
\EndWhile
\end{algorithmic}
\end{algorithm}

\emiAres{
We determine the success of the jamming attack based on the \ac{rssi}. This metric, unlike other metrics such as the \ac{per} or \ac{pdr}, does not require the attacker to spend a lot of time in listening mode. Therefore, the attacker can remain active for the whole duration of the attack. Indeed, we have observed a drop in the RSSI when an attack takes place whether on the side of the transmitter or the attacker.  }

\section{Channel Hopping Model} \label{section-mitigation}
In this section, we focus on the victim's side. We first describe in Section~\ref{sec:detection} the detection method implemented by the victim to detect an attack. Then, we describe in Section~\ref{sec:hoppingStrat} the different channel hopping strategies.

\subsection{Detection Method}\label{sec:detection}
To estimate the impact of FOLPETTI, we use the \ac{pdr}. This type of metric is often used to identify jamming attacks in \ale{the} literature \cite{metric-pdr}, and can be computed as
\begin{equation}
\label{pdr}
\mbox{Packet Delivery Ratio} = \frac{\sum \mbox{Number of PSD}}{\sum \mbox{Number of PT}},
\end{equation}
where \(PSD\) is the number of packets successfully received at the destination and \(PT\) represents the number of packets transmitted by the source.
\ale{Based on this metric, we implement a statistical detection framework based on the behavior of the network without attack~\cite{detection-method}.} Indeed, on the basis of a network without attack, it is possible to calculate the average PDR and thus to define a detection threshold $\delta$. If the PDR \ale{is smaller than} $\delta$, then the channel is under attack and the mitigation method can start. The choice of PDR as detection approach is mostly based on the consideration that it allows to detect different types of jamming attacks without increasing the computation overhead. 
\subsection{Channel Hopping}\label{sec:hoppingStrat}
Channel hopping is both an interference mitigation method and a reaction strategy against jamming attacks. \ale{It consists in dynamically changing the communication channel to mitigate interference or counteract to jamming attacks.} \ale{Assuming that the jammer does not attack on all the channels simultaneously, channel hopping is an effective method to keep the communication active.} However, there are some potential issues related to its implementation. \ale{For instance, the selection of the successive channel} should be realized without prior negotiation to avoid leakage of information that could be exploited by the attackers. Indeed, if the attacker eavesdrops the hopping pattern, it can continuously jam the network in all channels. \ale{In this work, we implement the following two strategies.}
\begin{itemize}
    \item \textbf{Random channel Hopping} $(Tx_{Random})$ : \ale{upon attack detection}, the transmitter node randomly chooses a new channel from the $M$ available ones. After the selection, it verifies the availability of the channel. If this channel is already occupied, it selects a new one. Therefore, the transmitter has a probability $M_{available} / M$ of selecting a \ale{free} channel, where $M_{available}$ represents the number of channel available at \ale{time} instant $t$.
    \item \textbf{Smart Channel Hopping} $(Tx_{Optimal})$: This method uses a \ac{mab} approach, and follows the model proposed by authors in~\cite{C-4}. This method tries to converge to the best channel available in as few steps as possible by employing a Thompson sampling formulation.
\end{itemize}

\section{Performance evaluation}\label{performance}
In this section, we first describe in Section~\ref{sec:networkModel} the system \ale{we consider} for implementing the attack based on the FOLPETTI strategy. \emi{Before} showing in Section~\ref{sec:performance} the performance obtained by running the FOLPETTI attack, \emi{we describe in Section~\ref{sec:otherStrat} the other channel  hopping strategies employed by the attacker}.

\subsection{Network Model}\label{sec:networkModel}
\begin{figure}[h]
\includegraphics[width=0.30\textwidth]{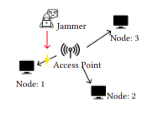}
\caption{\vspace{-0.5cm} Network model}
 \label{fig:network}
\end{figure}

We consider a network based on wireless nodes with limited energy, i.e. battery equipped. \ale{ All devices have similar features in terms of computational capacity and memory.} In particular, we consider a wireless communication network composed by three legitimate nodes connected via an access point with the IEEE 802.11 protocol and capable of transmitting on $12$ different channels~\cite{IEEE-ref}. We assume that the attacker has the same configuration as the legitimate nodes to reduce the probability of being detected. \ale{The attacker jams the communication between the access point and node 1 as depicted in Fig.~\ref{fig:network}.} We assume that legitimate nodes in the network may be able to detect an attack if their performance fall behind a certain value. Therefore, a stealthy attack is possible only if the attacker is able to keep its success rate below the identifiability threshold.

\ale{To follow the effects of the jamming attacks and compute the PDR,} the three considered devices communicate thanks to the Transmission Control Protocol (TCP). Indeed, TCP provides an acknowledgement (ACK packet) for each correctly received packet. Therefore, based on  these \ale{ACK} packets, the transmitter can judge  whether the transmission is successful and consequently update the PDR metric. We assume that the access point constantly transmits packets every $0.1$~s and begins its transmission at the start of the simulation ($t = 0$). After $10$~s, the attacker starts its attack. The legitimate nodes and the attacker start their communication on the same channel. Table \ref{parameters} summarizes the simulation parameters. 

\begin{table}[h]
\begin{center}
\begin{tabular}{|c|c|} 
 \hline
 \textbf{Parameter Name} & \textbf{Setting Used}   \\ 
 \hline
 Simulation Time & 1790 seconds  \\ 
 \hline
 Size of Legitimate Packet(octets) & 1000 \\
 \hline
 Start of jamming(seconds) & 10 \\
 \hline
  Start of channel hopping(seconds) & 1 \\
 \hline
 Energy Model & EnergyBasicModel \\
 \hline
 Distance Node - Access Point(m) & 5  \\
 \hline
  Threshold  Detection (\%) & 80 \\
  \hline
\end{tabular}
\caption{Simulation parameters}
\label{parameters}
\end{center}
\end{table}

\vspace{-0.5cm}
We \ale{implement} the aforementioned attacker's and transmitter's channel hopping methods in the discrete event simulator NS-3 (Network Simulator-3). We modify and update the jamming module~\cite{module-jamming} with the latest version of NS-3 and integrated into it the \textit{"ns3-gym"} module~\cite{ns3-gym} allowing to implement \ac{mab} algorithms. The code is available\footnote{If the reviewer wants to look at the code prior to publication, we can provide it through the program chairs. Due to the double-blind review process we cannot provide the link to the code at this stage.}.

\begin{figure}[h]
\includegraphics[width=0.40\textwidth]{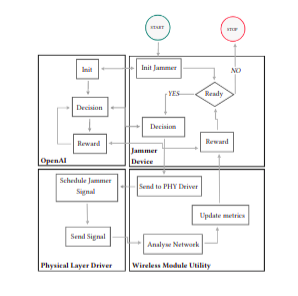}
\caption{\vspace{-0.5cm}  NS-3 Jammer work-flow}
 \label{fig:work-flow}
\end{figure}

Fig.~\ref{fig:work-flow} describes the work flow of the jamming attack we implemented in the NS-3 module. Four main components are required in the development of a jamming attack on NS-3. The first is the \textit{Jammer Device} class where the jammer strategy is defined. \textit{Wireless Module Utility} class is the second element of this model, This class provided essential functions for the jammer class such as functions to compute metrics of network performance. Moreover, this latter is also used to connect \textit{Jammer Device} and \textit{Physical Layer Driver} classes. Physical Layer Driver inherits from the physical class already implemented in NS-3. Thank to the latter we can define new behavior of the physical layer without modifying all the basic process of NS-3. The last part is the \textit{OpenAi} class which implements the logic of Multi-Armed Bandit algorithm. 

\smallbreak
Based on Fig.~\ref{fig:work-flow}, we describe the process of the FOLPETTI attack combined with the constant Jamming attack. The first step is to initialize the jammer attributes \ale{and the \ac{mab} parameters, i.e.}, the number of arms (in this case this corresponds to the number of channels). Once this step is done, the jammer is ready and has to make the decision on which channel to jam. To this aim, we exploit the \textit{OpenAi} class to return the arm chosen by the algorithm. Then, the \textit{Jammer Device} class sends all the necessary data to the \textit{Wireless Module Utility} class that transmits this information to the \textit{Physical Layer Driver} element. \ale{The Physical Layer Driver schedules} the next Jammer Signal. Once the signal has been sent, the \textit{Physical Layer Driver} notifies the \textit{Wireless Module Utility} which will then perform a network analysis. Then it returns an update of the different metrics of interest for FOLPETTI, i.e., the RSSI and the PDR. These metrics are communicated to the \textit{Jammer Device} class. Then, the \textit{Jammer Device} class receives the RSSI value and determines the reward. This value is delivered to the \textit{OpenAi} class to complete the procedure of the \ac{mab} algorithm. 

\smallbreak
\ale{ We developed this module to be as extensible as possible. Indeed, it is possible to extend all these classes to develop new jamming strategies, metrics, and new \ac{mab} algorithms. }

\subsection{Comparison with Other Attack Strategies }\label{sec:otherStrat}

\emi{In order to assess the validity of FOPLETTI, we compare it with four other attack strategies: }
%In the following sections, we evaluate our approach against two other attack strategies: 
\begin{itemize}
    \item \textbf{Random Channel Hopping} $(A_{Random}$): As for the mitigation method, the attacker randomly selects the channel on which to perform its attack. The channel change takes place after each transmission of the jamming signal. Hence in this case, the attacker has a probability of $1/M$ of jamming the correct channel, where $M$ represents the total number of channels.
    \emiAres{\item \textbf{Reactive Channel Hopping} $(A_{Reactive}$): One of the most developed strategies in the literature is the jamming reactive attack, which limits the attacker's energy consumption. The attacker jams the signal only when a communication takes place. This implies that the attacker must be able to react. The reaction time must be shorter than the packet transmission time for the latter to be an effect. Moreover, when the attacker senses no communication on a certain channel, he scans all the other channels in order to know the correct transmission channel.}
    
    \emi{\item \textbf{DRL based Channel Hopping}$(A_{DRL})$: The attacker employs the strategy developed in~\cite{zhong2020adversarial} and uses a \ac{drl} algorithm consisting of an actor and a critic to make decisions on the channel to attack. In this method, the actor observes its environment to choose the action to optimize his policy. During this time, the critic evaluates the actions performed by the actor by calculating the difference between the expected outcome and the actual one and informs it of the quality of its choice. The temporal difference value is then used to update the actor and critic model. Moreover, to optimize this model and reduce the probability of being detected, the attacker alternates between two modes: attacking phase, and listening mode. Indeed, the two agents (actor and critic) are based on neural networks that must be trained beforehand. This is why, during the training phase, the attacker is in listening mode. Once the model is trained, the attacker switches to attack mode. In this phase, the attacker scrambles the channel and can decide to switch to listening mode to re-train its neural network when performance decrease.}
    \item \textbf{Optimal Channel Hopping} $(A_{Optimal})$: We assume the attacker is omniscient and knows the victim's channel hopping pattern. After each diffusion of the jamming signal, the attacker verifies if it has to change channel. This approach represents the optimal solution, and we considered it as baseline in our simulations.
\end{itemize}

\subsection{Performance of FOLPETTI-Based Jamming }\label{sec:performance}
%To better compare and understand the results obtained with different attacking strategies, we initially describe the behavior of FOLPETTI against a random channel hopping strategy. 

\emi{To better understand the advantage of FOLPETTI, we initially describe its behavior and compare it with that of a  DRL-based attack. In this part, we assume that the victim's channel hopping strategy follows a random choice. }
\begin{figure}[h]
\includegraphics[width=0.40\textwidth]{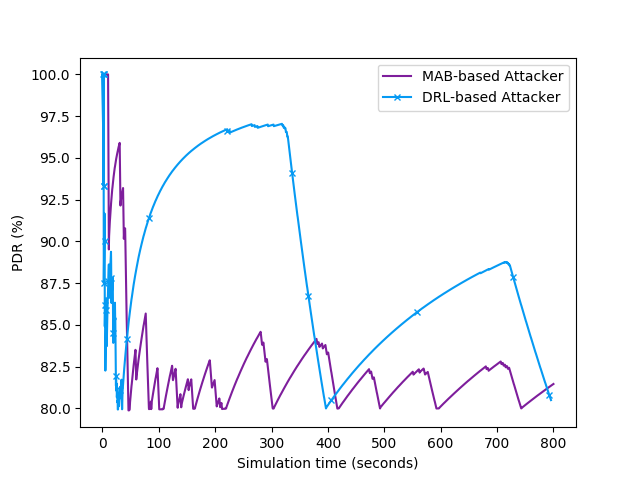}
\caption{\vspace{-0.5cm} \ac{pdr}  attacker.}
 \label{fig:pdr-one}
\end{figure}

\begin{figure}
     \centering
     \begin{subfigure}[b]{0.35\textwidth}
         \centering
         \includegraphics[width=\textwidth]{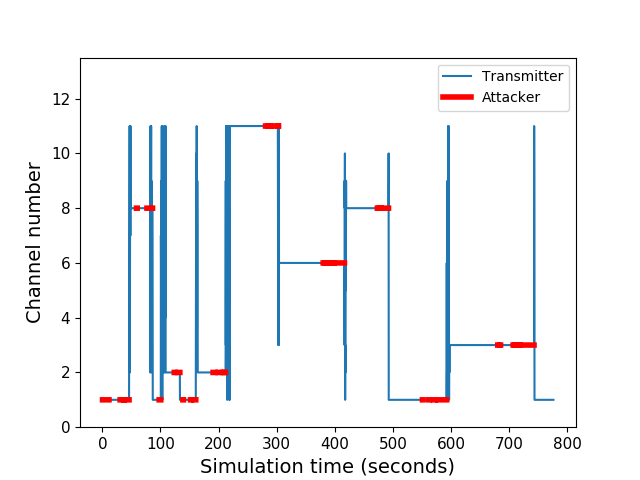}
         \caption{Channel Behavior with FOLPETTI attack}
         \label{fig:channel-png-th}
     \end{subfigure}
     \hfill
     \begin{subfigure}[b]{0.35\textwidth}
         \centering
         \includegraphics[width=\textwidth]{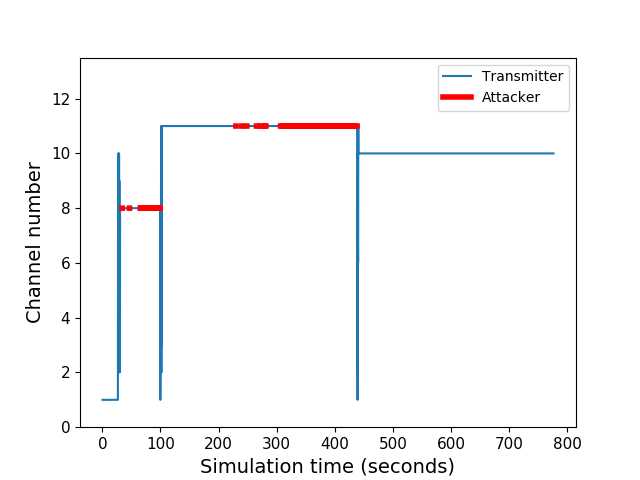}
         \caption{Channel Behavior with DRL-based attack}
         \label{fig:channel-png-drl}
     \end{subfigure}
        \caption{Transition channel pattern for transmitter and attacker \vspace{-0.2cm} }
        \label{fig:pattern-channel}
\end{figure}

\ale{Fig.~\ref{fig:pdr-one} shows the impact of FOLPETTI and DRL-based strategy on the PDR. The closer the PDR is to zero, the more effect the attack has on the network. We fix the attack detection threshold to $80\%$ PDR.} \emi{Indeed, based on \cite{Why-PDR}, and the distance between the access point and the victim that we defined above, the percentage of the PDR observed in real-life in a network without attack varies between 82\%- 100\% in a normal behavior.} Consequently, when the PDR drops at $80\%$ \emi{a jamming attack occurs and} the transmitter randomly hops into another channel. Fig.~\ref{fig:pattern-channel} shows the transition channel pattern for the transmitter and the attacker \emi{for the case of FOLPETTI attack and DRL-based attack}. For reasons of readability, we report the attacker's channel only when it is simultaneously used by the transmitter. 

\ale{By combining Fig.~\ref{fig:pdr-one} and Fig.~\ref{fig:channel-png-th}, we see that our attack tracks the channel hopping mitigation method.} Indeed, at $t = 10$ the PDR \ale{drops} from $100\%$ to $89\%$. This is explained by the fact that the attacker and the transmitter are positioned in the same channel. Then the attacker is in the exploration step and examines the availability of the other channels, hence increasing the PDR again to $95\%$. At the end of this phase, the attacker decides to turn to the exploitation period and jams channel 1 causing the PDR to drop to $80\%$. \ale{At this point, the transmitter detects a potential attack and changes the communication medium.} The victim randomly chooses a channel and verifies in a second time if this is occupied. Therefore at time $t=49$, the transmitter  switches to channel 11. However, as this has a low RSSI value, the access point will re-select a new channel (in this case, 8). Simultaneously, the attacker receives negative rewards and decides to change the jamming channel. Consequently, at $t = 54$ the attacker chooses to jam channel 8 which will again cause the PDR to drop to $80\%$.  \ale{We observe that this behavior pattern remains until the end of the simulation. Therefore, our proposed approach allows to quickly deduce the new communication parameters used by the transmitter.} Indeed, after the transmitter changes channel, the attacker can find the optimal channel so that the PDR does not exceed $84\%$.

\emi{On the contrary, for the DRL-based attack, we notice in Fig.~\ref{fig:pdr-one} that at the beginning of the simulation, the PDR rapidly drops to $80\%$. Consequently, the legitimate nodes of the networks react and change their transmission channel from 8 to 11, as we see in Fig.~\ref{fig:channel-png-drl}. However, the attacker continues to jam channel 8 for a short time until it finds that it has no effect on it. As a result, it switches to listening mode to re-train its neural network with new observations. This period appears on the figure starting from second $150$ and lasts about $100$~s. During this time, the attacker does not impact the network, and the values of the PDR increase.  At second $300$, the attacker finds its new policy and jams the channel where the transmission is taking place. Hence, the PDR values drop to the detection threshold and the transmitter will react accordingly by changing channel again. We note that this behavior is repeated until the end of the simulation. }

\emi{By analyzing the behavior of the two attacks, we see that the attack based on the \ac{mab} algorithm, i.e., FOLPETTI, reacts faster to policy updates than the one based on the DRL. This is confirmed by an overall lower \ac{pdr} for FOLPETTI. Indeed, this mechanism does not require any learning time, so the attacker can remain active throughout the attack, hence steadily keeping a close-to-optimal \ac{pdr}. }

\section{ASSESSMENT OF THE FOLPETTI ATTACK IN TWO SCENARIOS} \label{sec:evaluation}
\emi{In this section, we evaluate our new strategy in two scenarios. In the first Section~\ref{sec:scenario-1}, we consider a victim using random channel hopping. Then, we consider a victim exploiting a smart channel-hopping strategy in Section~\ref{sec:scenario-2}.}
For these two cases, we compare the different strategies in terms of a) PDR, b) success rate of the attack,  c) number of retransmissions and d) number of detections. The PDR is computed via~\eqref{pdr}. The success rate of the attack corresponds to the number of successfully jammed packets by the attacker over the total number of transmitted packets by the victim. The results obtained below correspond to an average over $1000$ simulations.

\subsection{Scenario 1: FOLPETTI Against Random Channel Hopping  } \label{sec:scenario-1}
\begin{figure}[h]
\includegraphics[width=0.40\textwidth]{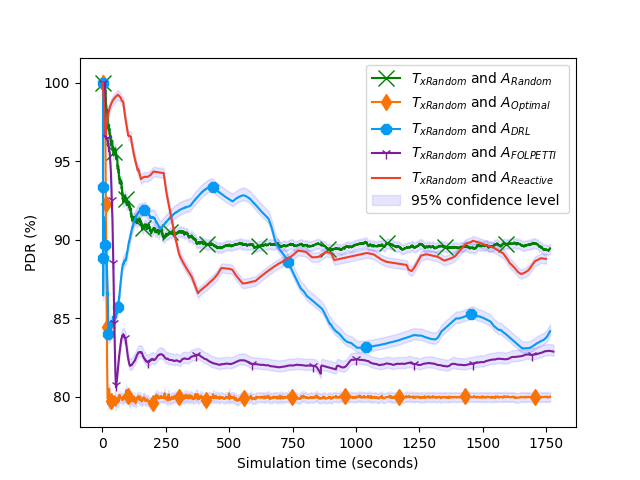}
\caption{ \ac{pdr} for different strategies of attack against Random Channel Hopping Method \vspace{-0.3cm}}
 \label{fig:ratio-random}
\end{figure}

%As we can see in the Fig.~\ref{fig:ratio-random}, for the FOLPETTI attack, the PDR  is close to that of the optimal solution. Indeed, the PDR after the beginning of the optimal attack is around $80\%$ against $82\%$ for our new approach. For the first type of attack, this is explained by the fact that the detection threshold is fixed at $80\%$. When the PDR is less than or equal to the threshold, the transmitter selects another channel. However, in this case, the attacker is all-knowing and therefore knows the future channel of the legitimate nodes. Hence, in this situation, the PDR is close to the detection threshold when an attack takes place. On the contrary, for the attack following the strategy based on random channel hopping, the value of the PDR remains high at $90\%$. This strategy is not effective because the attacker has a $1/13$ probability of jamming the correct channel. Consequently, the attack is ineffective and the PDR remains high.  
%However, our approach tries to get closer to the optimal solution with a PDR at 82\%.  

\emi{Fig.~\ref{fig:ratio-random} shows the impact of the different attacks on the network considering the PDR evolution over time. As can be seen in this figure, for the FOLPETTI attack the PDR values are the closest to those of the optimal solution. Indeed, the average PDR of the optimal solution is $80\%$ against $82\%$ for our new approach and $85\%$ for the DRL-based attack. It is important to note that the channel change by the transmitter takes place when the PDR is below a threshold  $\delta$, set here at $80\%$. In the case of the optimal solution, the attacker is omniscient and knows in advance the future channel to jam. Therefore, once the PDR reaches the detection threshold, it will remain constant throughout the simulation. Conversely, for the attack following the strategy based on random channel hopping, the value of the PDR remains high at $90\%$. This can be explained by the fact that the attacker has a $1/12$ probability of jamming the busy channel, therefore having limited effect on the transmission. \emiAres{The reactive attack has an efficiency close to the attack using a random channel strategy. Indeed, after 1 second of inactivity, the attacker re-scans all of the different channels in order to deduce the new transmission channel. Consequently, the attacker is inactive during this time and loses performance.} Finally, the DRL-based approach has an average PDR value of $85\%$ which is higher than of the FOLPETTI attack. In addition, the first learning time is considerable and the attack only manages to deduce a strategy after 600 s. In this approach, when the attacker considers itself ineffective, it has the possibility of re-training its two agents in order to re-adapt their strategies. As seen previously, during this phase, it cannot impact the network therefore having no effect on the PDR. On the other hand, the FOLPETTI attack has the ability to learn online and to remain active throughout the attack. Therefore the PDR obtained via FOLPETTI quickly decreases and closely approaches that of the optimal solution.}

\begin{figure}[h]
\includegraphics[width=0.40\textwidth]{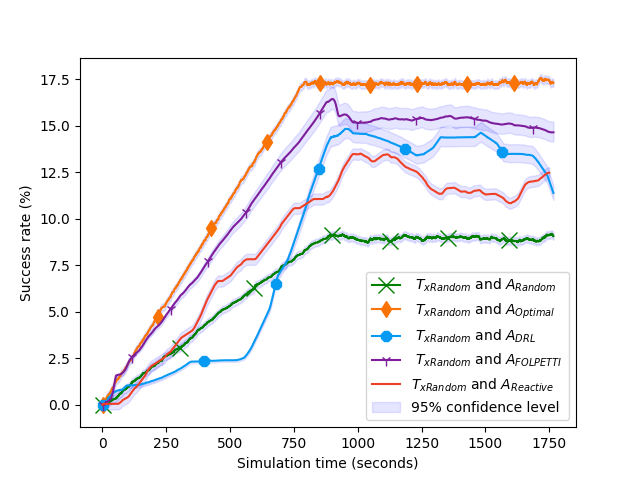}
\caption{Success rate for different strategies of attack against Random Channel Hopping Method}
 \label{fig:ratio-sucess}
\end{figure}

The performances of the different attack strategies evaluated according to the success rate are presented in Fig.~\ref{fig:ratio-sucess}. When the attack is optimal at the end of the simulation, the success rate is around $17.5\%$ against $15.5\%$ for the FOLPETTI solution \emi{and 12.5\% for the DRL-based solution}. 
%We also notice that FOLPETTI is almost twice efficient compared to the random solution. Indeed, the latter has a success rate of $8\%$. 
\emi{The FOLPETTI attack is more efficient than the other attack strategies and, in particular, twice as effective as that based on a random strategy. The difference between the FOLPETTI attack and the DRL-based attack can be explained by the fact that, in the first case, the attack needs fewer observations and therefore less time to converge towards the optimal solution.}

\begin{table}[h]
\begin{center}
\begin{tabular}{|p{3cm}|p{2cm}|p{2cm}|} 
 \hline
 \hfil \textbf{  Attack} & \textbf{Number of Retransmissions}  & \textbf{Number of detection}  \\ 
 \hline
 Random & 74 & 0 \\
  \hline
 Reactive & 419 & 514 \\
 \hline
 DRL-based Attack & 641 & 1189\\
 \hline
FOLPETTI & 738 & 1483 \\
 \hline
 Optimal & 1070 & 1729 \\
 \hline
\end{tabular}
\end{center}
\caption{ Number of retransmissions and detection for different attack strategies against random channel hopping method \vspace{-0.3cm}}
\label{scenario-1}
\end{table}

By comparing the number of retransmitted packets in Table~\ref{scenario-1}, we can discern that the strategy based on \ac{mab} with Thompson policy has an impact on the behavior of the network. On contrary, the random channel hopping strategy has no effect on the number of retransmissions. Indeed, the number of retransmissions is equal to $74$ against $738$ with the strategy based on the \ac{mab} and $1070$ for the optimal solution. \emi{Moreover, the number of retransmissions for the DRL-based solution is $641$, $13\%$ less than for the FOLPETTI attack.} 
The number of detection and consequently the number of channel hops at the transmitter side confirm these results. For the random solution, the PDR never drops below $80\%$, therefore no attack is detected and no channel hopping is performed. For FOLPETTI, the transmitter detects an attack $1483$ times against $1729$ times for the optimal solution. Therefore, this new type of approach produces a significant effect in terms of retransmissions and disturbance on the channel, hence decreasing the energy efficiency of the victim network.
\emi{As confirmed by the results obtained previously, the attack based on \ac{drl} is less efficient than the optimal and FOLPETTI attacks and therefore leads to a lower number of detection. Indeed, this approach has a detection number of $1189$, $294$ detections less than with our approach.}

Consequently, when the mitigation method is based on a random strategy, the proposed solution is effective and the results obtained are almost similar to those provided by an optimal solution.

\subsection{Scenario 2: FOLPETTI against Smart Channel Hopping} \label{sec:scenario-2}

In order to estimate the robustness of the FOLPETTI model, we evaluated it against a smart channel hopping method. We exploit the channel hopping method proposed in~\ref{section-mitigation}, that tries to predict the best communication channel available without interference. 

\begin{figure}[h]
\includegraphics[width=0.40\textwidth]{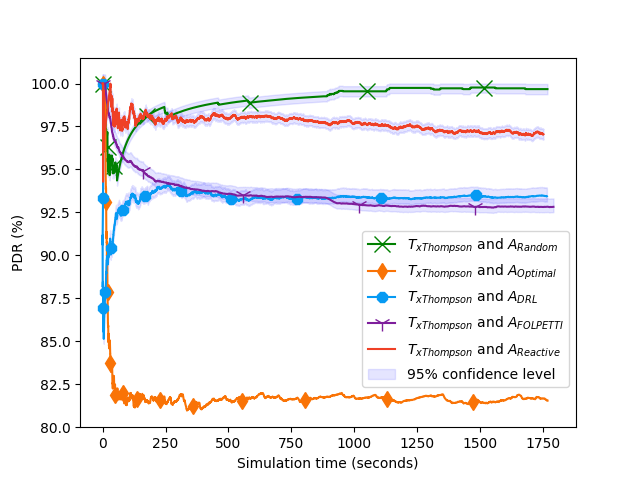}
\caption{\ac{pdr} for different strategies of attack against Smart Channel Hopping Method}
 \label{fig:ratio-pdr-th}
\end{figure}

The results obtained in Fig.~\ref{fig:ratio-pdr-th} reveal different points. First, the smart method obtains better performance in terms of defense compared to a random defense strategy. In this case, the PDR is equal to $100\%$ unlike $90\%$ when the channel hopping strategy is based on random selection. \emiAres{The same behavior is also visible for the reactive attack. Indeed, the PDR remains high around 97.5\%. }
\emi{For the two smart attackers, we notice that at the beginning of the simulation the PDR is around $94\%$. Therefore, this new types of attack decrease the PDR by $6\%$ compared to a more basic method. As our attack does not need training time and converges faster towards the choice of the optimal channel, the PDR tends to drop little by little to arrive at the end of the simulation at $92.5\%$. Thus, we notice that our approach has an impact on the network even if the mitigation method is more efficient. This is different from the attack based on the DRL algorithm, which at convergence reaches an average PDR value rise around $94.5\%$ due to the time needed to re-train the model.} 

\begin{figure}
\includegraphics[width=0.40\textwidth]{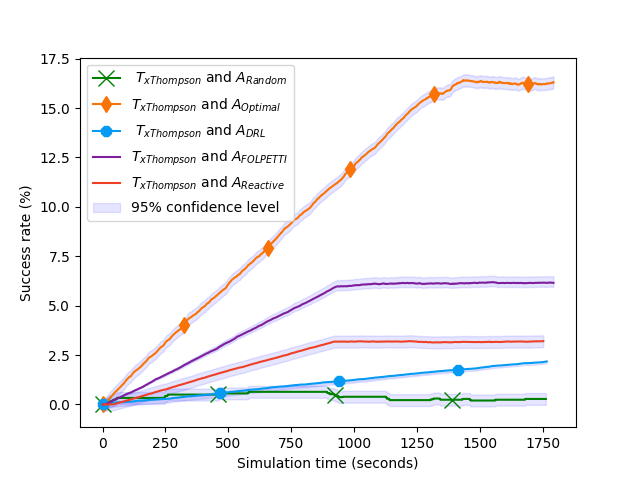}
\caption{Success rate for different strategies of attack against Smart Channel Hopping Method}
 \label{fig:ratio-sucess-th}
\end{figure}

These results are also confirmed by the success rate, shown in Fig.~\ref{fig:ratio-sucess-th}. Indeed, our new attack has a success rate $7.5\%$ higher than the one based on random channel hopping strategy, whose success rate is $0\%$. \emi{Moreover, if we compare the success rate between the two smart attacks, our approach has a higher success rate that converges around $7.5\%$ after $900$~s of simulation. During the same simulation time, the DRL-based attack has a success rate of $2\%$, i.e., $3.75$ times lower than that of the FOLPETTI attack. }

By looking at the number of retransmissions reported in Table~\ref{scenario-2}, the attack based on \ac{mab} algorithm has an important effect on the network as it allows to generate $280$ retransmission. \emi{The attack based on agent-critic method causes 212 retransmissions, i.e., 68 less than FOLPETTI attack.  }
In addition, the legitimate nodes must change channel $513$ times against $485$ times when the attack is based on DRL method and $0$ times with a random strategy. 

\begin{table}[h]
\begin{center}
\begin{tabular}{|p{3cm}|p{2cm}|p{2cm}|} 
 \hline
 \hfil \textbf{Attack} & \textbf{Number of Retransmission} & \textbf{Number of detection}  \\ 
 \hline
 Random  &  38 & 0 \\
 \hline
  Reactive  & 87  & 124 \\
  \hline
  DRL-based Attack  &212 &485  \\
  \hline
  FOLPETTI  & 280 & 513 \\
 \hline
 Optimal & 470 & 751\\
 \hline
\end{tabular}
\end{center}
\caption{Number of retransmission and detection for different attack strategies against smart Channel Hopping method}
\label{scenario-2}
\end{table}

Simulations demonstrated the efficiency of this new type of attack to select the optimal channel to jam even if the channel hopping method is more robust. 

\vspace{-0.2cm}
\section{Discussion}\label{discussion}

Based on the results obtained in the two scenarios considered, we observe that the proposed method has better performance than a basic strategy like a random method. In addition, FOLPETTI is also effective when the channel hopping strategy is based on a more advanced approach such as the \ac{mab} algorithm. The main advantage of FOLPETTI based attack is that it operates without having prior knowledge of its victim. Indeed, the apprenticeship is completely online, the attacker does not alternate between long transmission and listening phases in order to discern its impact on the victim. This is possible thanks to the RSSI metric which does not require any listening time, allowing the attacker to remain active throughout its attack. Another main benefit of Multi Armed Bandit algorithm with Thompson policy is that it requires less computing resources and can be executed in small devices as evidenced in \cite{C-4}. Therefore, the attacker can use common devices whilst being mobile like a Raspberry Pi, and executes attacks in any environment. 

\smallbreak

By analyzing the number of retransmission that FOLPETTI generates, we can deduce that this type of attack strongly impacts on the network performance. An increase in the number of retransmissions not only causes a delay in the delivery of information but also a considerable loss of energy for the transmitter. Indeed, the energy consumption $E_{\text{total}}$ of the transmitter in 802.11 protocol can be computed with this formula: 

\begin{equation}
\begin{split}
E_{\text{total}} = PW_{Tx} \times T_{Tx} +  PW_{Rx} \times  T_{Rx} + \\
PW_{IDLE} \times  T_{IDLE} + PW_{SLEEP} \times  T_{SLEEP}, \\
\end{split}
\label{enegyequation}
\end{equation}

where $PW_{TX}$, $PW_{RX}$, $PW_{IDLE}$ and $PW_{SLEEP}$ represent the power consumption of the transmitter node when it is in the transmitting, listening, idle, and sleep states respectively. $T_{TX}$, $T_{RX} $, $T_{IDLE}$ and $T_{SLEEP}$ correspond to the total time spent by the transmitter in each respective state.
In addition, the total of energy for one retransmission $E_{Retx}$ can be described as: 

\begin{equation}
    E_{Retx} = R ( T_{Tx}  \times PW_{Tx}) + R ( T_{Rx} \times PW_{Rx}),
    \label{energy-transmitter}
\end{equation}

where $R$ corresponds to the total number of packets transmitted during one retransmission such as data or acknowledgements packets. 
Diverse measures of power consumption for several type of network interface controller are provided in \cite{energy}. The authors show that the transmitting and listening are the two most energy demanding modes. Combining this consideration with equations \eqref{enegyequation} and \eqref{energy-transmitter}, we notice that FOLPETTI drastically reduces the lifetime of a device. Indeed, if the attacker is equipped with an Alfa AWUS036h network interface controller compatible with a Raspberry Pi, the energy consumption for each step can be determined as shown in table \ref{tab:energy}.  These information are provided by \cite{atheros1}.

\vspace{-0.2cm}

\begin{table} [h]
\begin{center}
\begin{tabular}{ |c|c|c|c|c|} 
 \hline 
  \textbf{Operating Mode} & \textbf{Sleep}& \textbf{Idle} & \textbf{Tx} & \textbf{Rx} \\
   \hline 
   \textbf{P(W)} &   0.001 &   0.30 &   0,67 &  0,34\\ 
\hline
\end{tabular}
\caption{Power Consumption for 2.4 GHz Operation}
\label{tab:energy}
\end{center}
\end{table}

\vspace{-0.7cm}

Based on the previous simulations and by considering the values given in Table \ref{tab:time} corresponding  to the transmission $T_{Tx}$ time and reception time $T_{Rx}$ of the various packets involved for one retransmission, we can compute the additional network energy consumption under a jamming attack. 

\begin{table} [h]
\begin{center}
\begin{tabular}{ |c|c|c|} 
 \hline 
  \textbf{Type of packet} & \textbf{Transmission time (s)} & \textbf{Reception time (s)}\\
   \hline 
   \textbf{Data} & 0,00397 & 0,00695 \\ \hline
   \textbf{Ack} &  0,00002 & 0,00003\\ \hline
\end{tabular}
\caption{Average time spent by each state during a retransmission according to the type of packet}
\label{tab:time}
\end{center}
\end{table}
\vspace{-0.7cm}
Combining formula~\eqref{energy-transmitter} and the information from Table~\ref{tab:energy}, we deduce that the energy consumption for one retransmission $E_{Retx}$ is $0.0050431$~J. For the scenario 1, the number of retransmission is 738 when the strategy attack is based on Multi Armed Bandit. Therefore the additional energy expended by the transmitter is $3.72$~J for an attack of 30 minutes. 
\emi{For the same situation, the DRL-based attack generates an increase of energy consumption of $3.23$~J, which is $0.49$~J less than the FOLPETTI attack.} With the random jamming approach, the supplementary energy consumed by the access point is  $0.373$~J, i.e 10 times less than for FOLPETTI method.

The same results are also observed for the second scenario.  Indeed, for 30 minutes of simulations, the number of retransmissions is $280$ when the jammer is based on the new approach and $38$ when the attacker applies a random logic. Consequently, the emitter consumes $1.41$~J with the FOLPETTI model against $0.19$ J with a basic method. \emi{Our new approach has a greater effect on the victim's energy consumption than other attack strategies. This is confirmed comparing our attack to the DRL-based. Indeed, the \ac{drl}-based method causes an energy expenditure for the victim of $1.06$~J, i.e. $0.35$~J less than the FOLPETTI attack. }

\smallbreak

\emi{The main advantage of FOLPETTI attack is that it does not require training. Indeed, the update of its policy is done in real time. Unlike other works, we do not rely on ACK packets to define the success of our attack. Consequently, the attacker must not wait to receive this packet to recover a reward. The update of the policy is done in a smaller time and the attacker will converge faster to the optimal solution. As the result, FOLPETTI attack is more effective as we proved in previous section. }

\smallbreak
To finish, our new type of attack can be considered as a framework to perform several known attacks. Indeed, being able to predict the communication channel can be useful in replay attack or flooding attack in order to limit their probability of detection. Moreover, more and more attacks are based on machine learning algorithms and consequently data sets in order to deduce information~\cite{ATPP-1,ATPP-5}. In particular, they require passive listening to the network in order to collect the maximum of data. This step is crucial because the effectiveness of the machine learning algorithm relies on the data set. In this case, FOLPETTI makes it possible to follow the channel hop and thus not to lose any information necessary for the creation of the data set. 
In addition, the mitigation method by channel hopping is not only present in Wi-Fi protocol. Indeed, this strategy is also employed by other communication protocols. Thereby, this new framework does not depend on the Wi-Fi protocol and can be effective in other use cases. 

%such as the 802.15.4 TSCH Networks~\cite{802.15} or Bluetooth Network~\cite{BLE}%

\section{Possible Countermeasures}\label{sec:countermeasures}

\emi{We have just shown that the different strategies employed in the channel hopping literature lose effectiveness when the attacker operates with a more elaborate algorithm. Indeed, even if the FOLPETTI attack has smaller effect when a smart channel hopping is used, this new type of attack leads to a decrease in network performance. Several works aimed at creating more robust \ac{mab} algorithms exist in the literature and could be applied in the \ac{mab} channel hopping strategies~\cite{CP-3}. Moreover, it is also possible to evade FOLPETTI attack by directly attacking the \ac{mab} model of the attacker. Indeed, this type of algorithm can be subjected to adversarial attacks, and two strategies could be designed to disturb the FOLPETTI model.
\begin{itemize}
    \item {\textbf{Poisoning attacks:}} One of the countermeasures can be the poisoning attack~\cite{CP-1}. The purpose of the transmitter is to deceive the attacker about its choice. The victim influences the attacker to jam an unused channel, having hence no effect. To this aim, legitimates nodes must falsify certain rewards in order to control the attacker's decision. In this situation, nodes can possibly voluntary transmit in an jammed channel to influence the number of rewards and therefore the attacker's policy. 
    \item{\textbf{Delayed feedback attack:}} Another possible countermeasure could be to implement delayed feedback in order to deceive the attacker's choice. The goal of the victim is to delay the transmission of feedback information that serves as a reward to the attacker.  In this case, the complexity of the \ac{mab} algorithm for the attacker will increase significantly as demonstrated in~\cite{CP-2}. As the result, the attacker will need a significant time to converge to the optimal solution.
\end{itemize}}

\emi{One of the main difficulties of these solutions is that the transmitter must completely know the parameters of the \ac{mab} algorithm employed by the attacker, such as the type of the rewards or the policy. In addition, the legitimates nodes of the network must be able to implement these methods without impacting the network performance. Finally, an important point to take into account when designing these countermeasures in the IoT context is the power consumption they require. Consequently, designing solutions against FOLPETTI attacks requiring few resources, knowledge, and not impacting the network remains an open challenge. }
\balance

\section{Conclusion and Future work}\label{concl}

In this article, we presented FOLPETTI, a novel smart attack that operates in an unknown network in the absence of a-priori information about the network itself. FOLPETTI has the ability to understand and predict the behaviour of victims and more particularly their channel hopping strategy. We evaluated FOLPETTI against two defence strategies, one more classic based on random channel hopping and one more advanced based upon \ac{mab} algorithm.  In the first situation, FOLPETTI approaches an optimal attack solution in terms of PDR, success rate, and number of retransmissions. In the second case, our new attack is still able to enhance the success rate to $5\%$ against $2.5\%$ for the other smart attack.  The simulations results demonstrated that the proposed strategy can improve the overall jamming performance. Moreover, our approach can be considered as a framework to be used for any multi-channel communication protocol. Although we used FOLPETTI to perform a jamming attack, it can also be used for other attacks that may benefit from channel following strategies such as replay attack.
Our future works will evaluate this new solution in a testbed composed of several nodes and jammers in order to evaluate its performance in real life situation. \emi{Moreover, since the final goal of creating an attack is to improve the security system, we will add countermeasures in the smart channel hopping method against a FOLPETTI attack. This study will focus on the trade-off between the security, efficiency, and the network performance. }

%Moreover, we will improve the convergence time, namely the time for learn and switch to the best channel, by including a real "memory" of decisions adopted upstream including a deep learning algorithm. The goal would be to compare the two approaches in terms of impact on the network but also of learning time and energy consumption necessary for the attacker. Indeed, this solution could reduce the convergence time, however, its impact can also be close to that explained in this paper, all the while consuming more energy. This study will, therefore, focus on the trade-off between the attacks overall performance and the quantity of resources required. 

%\section{Acknowledgments}

\bibliographystyle{ACM-Reference-Format}
\bibliography{bibliography}

\end{document}